\documentclass[%
aps,
prx,
twocolumn,
superscriptaddress,
amsmath,
nofootinbib,
reprint
]{revtex4-2}
\usepackage[separate-uncertainty=true,binary-units]{siunitx}
\usepackage{graphicx}
\usepackage{tikz}
\usepackage{ifthen}
\usepackage[utf8]{inputenc}
\usepackage{enumitem} 
\usepackage{amsmath, amssymb}
\usepackage{siunitx}
\usepackage{lipsum}
\usepackage{csquotes} 
\usepackage{paralist} 

\usepackage{pdfpages} 
\makeatletter
\AtBeginDocument{\let\LS@rot\@undefined}
\makeatother

\definecolor{nicegray}{HTML}{555555}
\definecolor{nicered}{HTML}{AF5A50}
\definecolor{niceblue}{HTML}{005B82}
\definecolor{nicegreen}{HTML}{7D966E}
\definecolor{niceyellow}{HTML}{D7AA50}

\newcommand{\lat}{\mathrm{lat}}
\newcommand{\ift}{\mathrm{inf}}
\newcommand{\potInfEnc}{n_\ift}
\newcommand{\pext}{p_\mathrm{ext}}

\begin{document}

\title{How contact patterns destabilize and modulate epidemic outbreaks}

\author{Johannes Zierenberg}
\email{johannes.zierenberg@ds.mpg.de}
\thanks{JZ and FPS contributed equally}
\author{F.~Paul Spitzner}
\thanks{JZ and FPS contributed equally}
\author{Jonas Dehning}
\affiliation{Max Planck Institute for Dynamics and Self-Organization, 37077 G\"ottingen, Germany}
\author{Viola Priesemann}
\affiliation{Max Planck Institute for Dynamics and Self-Organization, 37077 G\"ottingen, Germany}
\affiliation{\mbox{Institute for the Dynamics of Complex Systems, University of G\"ottingen, 37077 G\"ottingen, Germany}}
\author{Martin Weigel}
\affiliation{Institut f\"ur Physik, Technische Universit\"at Chemnitz, 09107 Chemnitz, Germany}
\author{Michael Wilczek}
\affiliation{Max Planck Institute for Dynamics and Self-Organization, 37077 G\"ottingen, Germany}
\affiliation{Theoretical Physics I, University of Bayreuth, 95440 Bayreuth, Germany}

\date{\today}
\begin{abstract}
The spread of a contagious disease clearly depends on when infected individuals come into contact with susceptible ones.
Such effects, however, have remained largely unexplored in the study of epidemic outbreaks.
In particular, it remains unclear how the timing of contacts interacts with the latent and infectious stages of the disease.
Here, we use real-world physical proximity data to study this interaction and find that the temporal statistics of actual human contact patterns i) destabilize epidemic outbreaks and ii) modulate the basic reproduction number.
We explain both observations by distinct aspects of the observed contact patterns.
On the one hand, we find the destabilization of outbreaks to be caused by temporal clustering of contacts leading to over-dispersed offspring distributions and increased probabilities of otherwise rare events (zero- and super-spreading).
Notably, our analysis enables us to disentangle  previously elusive sources of over-dispersion in empirical offspring distributions.
On the other hand, we find the modulation of $R_0$ to be caused by a periodically varying contact rate.
Both mechanisms are a direct consequence of the memory in contact behavior, and we showcase a generative process that reproduces these non-Markovian statistics.
Our results point to the importance of including non-Markovian contact timings into studies of epidemic outbreaks.
\end{abstract}

\maketitle

\section{Introduction}

As contagious diseases are passed on through contacts, the number of secondary infections depends crucially on the contact patterns of infectious individuals.
These contact patterns encode relevant information such as the number of interaction partners and contact timing.
However, the majority of prevailing models for disease spread prioritize simpler descriptions that neglect these aspects ---
despite evidence from studies that show the effects of contact patterns to be crucial for disease spread:
Structurally, when \emph{interaction partners} are modeled by a static complex network~\cite{pastor-satorras_epidemic_2015}, the network structure affects disease spread through the occurrence of hubs~\cite{newman_why_2003, goltsev_localization_2012}, multiscale link communities~\cite{ahn_link_2010} and influential spreaders~\cite{kitsak_identification_2010}.
Dynamically, real-world \emph{interaction times} generally follow a non-Markovian process (in contrast to commonly assumed memoryless processes), which influences epidemics through the occurrence of bursts~\cite{barabasi_origin_2005, van_mieghem_non-markovian_2013} and daily and weekly variations in human interaction~\cite{du_periodicity_2018, towers_impact_2012}.

Thus, for a better understanding of disease spread through human contacts, a complete description of time-varying interactions in the form of so-called \emph{temporal networks}~\cite{masuda_predicting_2013, holme_modern_2015} seems necessary.
However, constructing detailed temporal networks from real-world contacts requires extensive amounts of recorded data,
which in principle can be collected in field studies~\cite{sekara_fundamental_2016, genois_can_2018, schlosser_covid-19_2020} but such data are notoriously limited in duration and system size.
Although such limitations of real-world data can be partly remedied by generating surrogate data~\cite{leitch_toward_2019, presigny_building_2021}, it is often unclear to which extent they represent the real system.
An unsolved task is thus to generate surrogate data that mimics temporal statistics and individual variations of human contact data.


Here, we address this gap by identifying and isolating features of contact behaviour that affect epidemic outbreaks using a novel analysis of real-world contact data.
Instead of characterizing full epidemic outbreaks on a large (likely under-determined) temporal network, we develop an effective description through \textit{potentially infectious encounters} that propagates statistics of contacts to statistics of disease spread.
This approach avoids treating microscopic (non-linear) network effects~\cite{kiss_mathematics_2017, zierenberg_description_2020, nie_effects_2022} and allows us to focus on how contact patterns statistically affect epidemic outbreaks.
Our analysis reveals two main mechanisms:
(i) contact clustering destabilizes outbreaks by increasing the dispersion of offspring distributions and the probability of zero-spreading events, and
(ii) temporal variation of the contact rate modulates the mean basic reproduction number, $R_0$, due to an interference between contact patterns and disease progression.
Finally, we showcase a non-Markovian process that faithfully reproduces the temporal statistics and their effect on disease spread as a proof of principle for a new class of generative models for surrogate data that mimic human contact patterns.

%

\section{Results}
\begin{figure}[t!]
    \centerline{\includegraphics[]{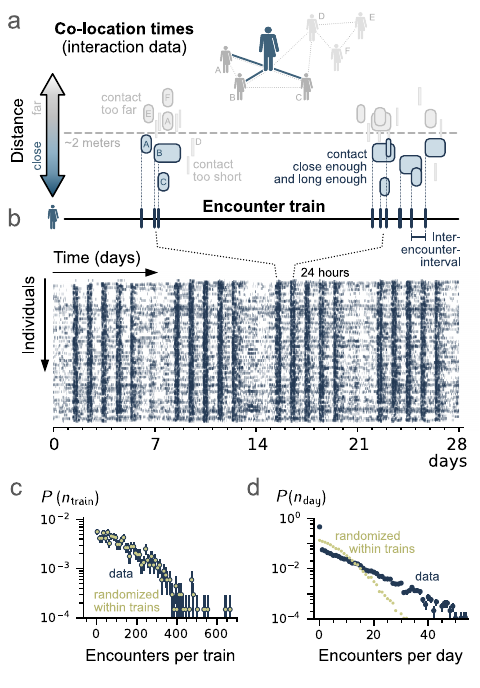}}
    \caption{
    \textbf{Real-world contacts represented as encounter trains.}
    \textbf{a:}~A contact between two individuals is defined as ongoing co-location during consecutive time steps.
    Focusing on contagious disease transmission, we only consider contacts closer than 2 meters and longer than 15 minutes.
    \textbf{b:}~The start times of the remaining contacts per individual form their encounter train.
    A raster plot of such encounter trains shows a clear temporal structure of human contact patterns.
    \textbf{c,~d:}~Randomization preserves the number of encounters per train but destroys temporal structure.
   }
    \label{fig1}
\end{figure}
To shed light on the interplay of contact patterns and epidemic outbreaks,
we analyse proximity data from the Copenhagen Networks Study~\cite{sapiezynski_interaction_2019} and from SocioPatterns~\cite{noauthor_sociopatterns_nodate}.
We filter each individuals' contacts by distance and duration, and define \emph{encounters} as their starting times (see Methods). The resulting \textit{encounter trains} are a point-process-like representation that captures the non-Markovian statistics of the underlying contact patterns (Fig.~\ref{fig1}).
The importance of these non-Markovian statistics can be seen when comparing to \textit{randomized encounter trains}.
In these surrogate data, encounter times are uniformly reassigned within the duration of the experiment --- which preserves the number of encounters per train, i.e., the inter-individual variability (Fig.~\ref{fig1}c), but destroys any temporal structure (Fig.~\ref{fig1}d).

In order to quantify the effect of contact patterns on epidemic outbreaks, we focus on {\em potential} secondary infections and assume a contagious disease that can be transmitted only during infectious encounters.
Further, if $\tau$ is the time elapsed since infection, infected individuals undergo a (non-infectious) latent period $\tau\in[0,T_\lat)$ and an infectious period $\tau\in[T_\lat,T_\lat+T_\ift)$ during which
2all $\potInfEnc$ encounters are \textit{potentially infectious}.
We estimate $\potInfEnc$ by considering every encounter in the data set as a potential start for an infection (Fig.~\ref{fig2}a, see Methods).
As we show in Fig.~\ref{fig2}b, empirical contact patterns increase both the probability of very few potentially infectious contacts $\potInfEnc$ (related to zero-spreading events) as well as very many $\potInfEnc$ (related to super-spreading events) when compared to randomized controls.
This increase in variability influences whether a single infection results in an epidemic outbreak or not.

\subsection*{Human contact patterns destabilize epidemic outbreaks}
\begin{figure*}[ht!]
    \centerline{\includegraphics[]{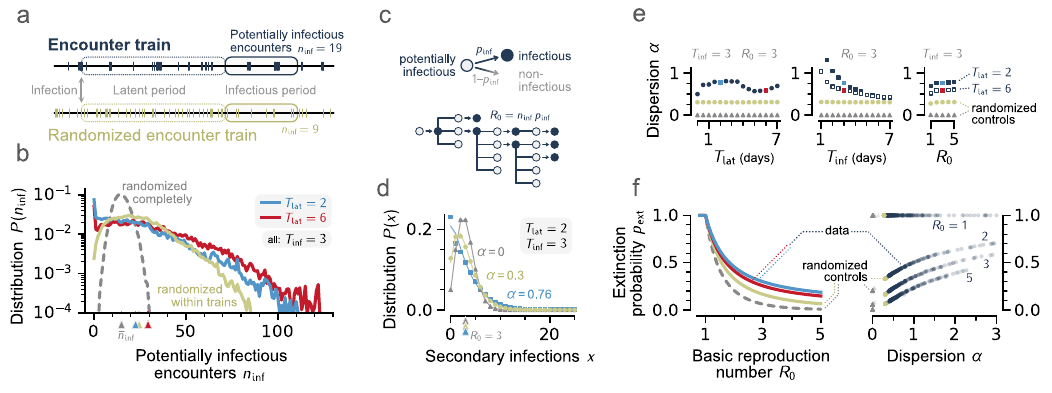}}
    \caption{
    \textbf{Real-world contact patterns increase the probability of rare spreading events, which increase the extinction probability of an epidemic outbreak.}
    \textbf{a:}~Sketch of estimating $n_\ift$ from encounter trains for a disease model with $T_\lat$ and $T_\ift$.
    \textbf{b:}~$P(n_\ift)$ depends on $T_\lat$ for human encounter trains (red, blue) but not for randomized controls (gray, yellow), which underestimate both zero-spreading as well as super-spreading events.
    \textbf{c:}~Data-driven branching model where $n_\ift$ are drawn from $P(n_\ift)$ and infected independently with probability $p_\ift$.
    \textbf{d:}~The resulting offspring distribution $P(x)$ can be fitted by a negative binomial, yielding an estimate for the dispersion parameter $\alpha$.
    \textbf{e:}~Dispersion as a function of $T_\lat$, $T_\ift$ and $R_{\rm 0}$.
    \textbf{f:}~The extinction probability of an epidemic outbreak depends on $R_{\rm 0}$ (left, fixed $T_{\inf} = \SI{3}{days}$) and on $\alpha$ (right, all combinations of $T_\lat$ and $T_\ift$ for given $R_0$), and is larger for human contact patterns compared to randomized controls.
   }
    \label{fig2}
\end{figure*}

To demonstrate the effect of empirical contact patterns on epidemic outbreaks, we map the probability distribution of $\potInfEnc$ to an offspring distribution using a two-step, data-driven branching process (Fig.~\ref{fig2}c):
Each infected individual first generates encounters according to the empirical $P(\potInfEnc)$, and then infects each of them independently with probability $p_\ift$ resulting in binomial-distributed secondary infections.
Taking the expectation value yields the offspring distribution
\begin{equation}
P(x)=\sum_{\potInfEnc=x}^\infty P(\potInfEnc)\binom{\potInfEnc}{x}p_\ift^x(1-p_\ift)^{\potInfEnc-x}.
\end{equation}
Similar to empirical distributions from contact tracing~\cite{lloyd-smith_maximum_2007}, $P(x)$ can be well described by a negative binomial distribution (Fig.~\ref{fig2}d) with mean $\overline{x} = R_0 = p_\ift\,\overline{n}_\ift$ and variance $\overline{(x-R_0)^2} = R_0 + \alpha R_0^2$, where $\alpha$ is the dispersion parameter that characterizes the increase in variance relative to a Poisson distribution.

Our data analysis provides a systematic approach to identify sources of the dispersion observed in empirical offspring distributions~\cite{lloyd-smith_superspreading_2005,peak_comparing_2017,hellewell_feasibility_2020}.
We analyze step by step the dispersion occurring because of human contact patterns, and how it depends on $T_\lat$, $T_\ift$ and $R_0$ (Fig.~\ref{fig2}e, left to right).
For a completely randomized control, where encounters are uniformly distributed across trains and time, we consistently find Poissonian offspring distributions (Fig.~\ref{fig2}d, gray), with low dispersion ($\alpha=0$), independent of the three disease parameters (Fig.~\ref{fig2}e, gray triangles).
When including variability of contact rates into the control, while still randomizing within trains, the dispersion of the offspring distribution increases ($\alpha\approx0.3$) but remains mostly independent of disease parameters (yellow circles).
Lastly, when also including the precise timing of human contact patterns, offspring distributions show large dispersion that depends on $T_\lat$ and $T_\ift$ (blue symbols). In particular, dispersion is strongest for short $T_\ift$ but decays as $T_\ift$ increases.
Hence, part of the empirical dispersion can be attributed to variability of contact rates between individuals, but non-Markovian timing of human contact patterns causes a severe increase --- for realistic parameters roughly by a factor of two.

From $P(x)$ we derive the extinction probability $\pext$, defined as the fraction of outbreaks that asymptotically end up in the absorbing state of zero infections (Fig.~\ref{fig2}f).
It can be calculated using the probability generating function, $\pi(\theta)=\sum_{x=0}^\infty P(x)\theta^x$, as the smallest $\theta^\ast$ that solves $\theta^\ast=\pi(\theta^\ast)$~\cite{harris_theory_1963}.
In addition to the anticipated monotonic decrease of $\pext=\theta^\ast$ with increasing $R_0$, we find that extinction is more likely for actual human contact patterns (red, blue) than for randomized controls (gray, yellow).
Moreover, for fixed $R_0$, we find that an increased dispersion $\alpha$ due to human contact patterns non-linearly increases $\pext$ (Fig.~\ref{fig2}f, right).

Summarizing, we find that the non-Markovian timing of human contact patterns can be a strong source of variability, relevant to explain the over-dispersion of empirical offspring distributions. In particular, increasing the dispersion for a fixed $R_0$ increases the probability of zero-spreading events (Fig.~\ref{fig2}d, blue vs gray), and results in a higher extinction probability (Fig.~\ref{fig2}f) --- in other words, the non-Markovian temporal structure of human contact patterns destabilizes epidemic outbreaks.

\begin{figure*}[hbt!]
    \centerline{\includegraphics[]{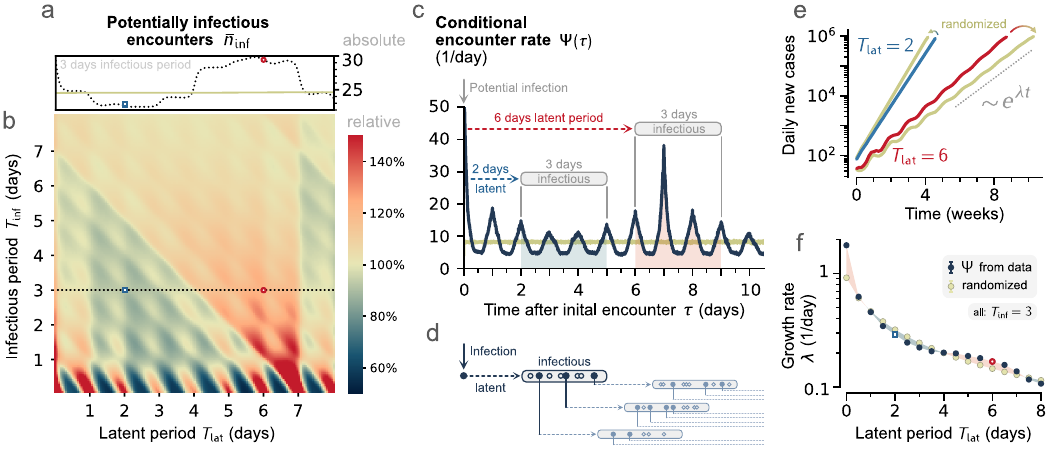}}

    \caption{
    \textbf{Real-world contact patterns modulate the pace of epidemic spread as a function of latent period.}
    \textbf{a:}~Absolute $\overline{n}_\ift$ are periodically modulated with $T_\lat$ for human encounter trains but not for randomized (yellow).
    \textbf{b:}~Relative $\overline{n}_\ift$/$\overline{n}_\ift^\mathrm{rand}$ reveals periodic modulations on both daily and weekly scale in the full $(T_\lat,T_\ift)$-plane.
    \textbf{c:}~$\Psi(\tau)$
    features daily and weekly modulations for non-Markovian human encounter trains but is constant for randomized encounter trains (yellow).
    Assuming the initial encounter to be an infection, this explains the modulations of $\overline{n}_\ift$ by combinations of $T_\lat$ and $T_\ift$ for which the integral (shaded areas) is dominated by valleys (blue) or peaks (red).
    \textbf{d:}~Continuous-time branching model, where encounter times are generated from $\Psi(\tau)$ and infected with constant probability $p_\ift$.
    \textbf{e:}~Starting from an initial $I_0=100$ random infections in $[-T_\lat-T_\ift,0)$,  the (average) number of infections grows exponentially.
    The growth rate $\lambda$ for time-independent encounter times in a fixed $T_\ift$ is expected to decrease trivially with $T_\lat$.
    If not constant, $\Psi(\tau)$ modulates $\lambda$ and causes regimes of slower-than-random (blue) or faster-than-random (red) growth of infections.
    }
    \label{fig3}
\end{figure*}

\subsection*{Interplay between contact pattern and disease progression modulates basic reproduction number}

As highlighted in Fig.~\ref{fig2}b,
$\overline{n}_\ift$ depends on $T_\lat$ for human encounter trains.
This is at first glance surprising, because for memory-less processes $\overline{n}_\ift$ is proportional to $T_\ift$ but independent of $T_\lat$.
Hence, in the following, we systematically vary $T_\lat$ and $T_\ift$ to study how the interplay between human contact patterns and disease progression affects $\overline{n}_\ift$ (Fig.~\ref{fig3}).

Considering a fixed $T_\ift=\SI{3}{days}$ (Fig.~\ref{fig3}a) and scanning $T_\lat$ leads to a periodic modulation of $\overline{n}_\ift$ from human encounter trains (black, dashed) around the constant estimate from randomized trains (yellow).
Thus, we consider $\overline{n}_\ift$ relative to randomized (Fig.~\ref{fig3}b), which accounts for the trivial increase of $\overline{n}_\ift$ with increasing $T_\ift$.
For small $T_\ift<\SI{1}{day}$, we find daily modulations as a function of $T_\lat$, with regions below-randomized (blue) and above-randomized (red).
This effect diminishes for larger $T_\ift$, where we find extended, triangular regions with interfaces located at $T_{\rm lat}+T_{\rm inf}=\SI{7}{days}$ and $T_\lat=\SI{7}{days}$.
We thus find periodic modulations of $\overline{n}_\ift$ on the scale of days (small $T_\lat$) and weeks (large $T_\lat$).

\begin{figure*}[t]
    \centerline{\includegraphics[]{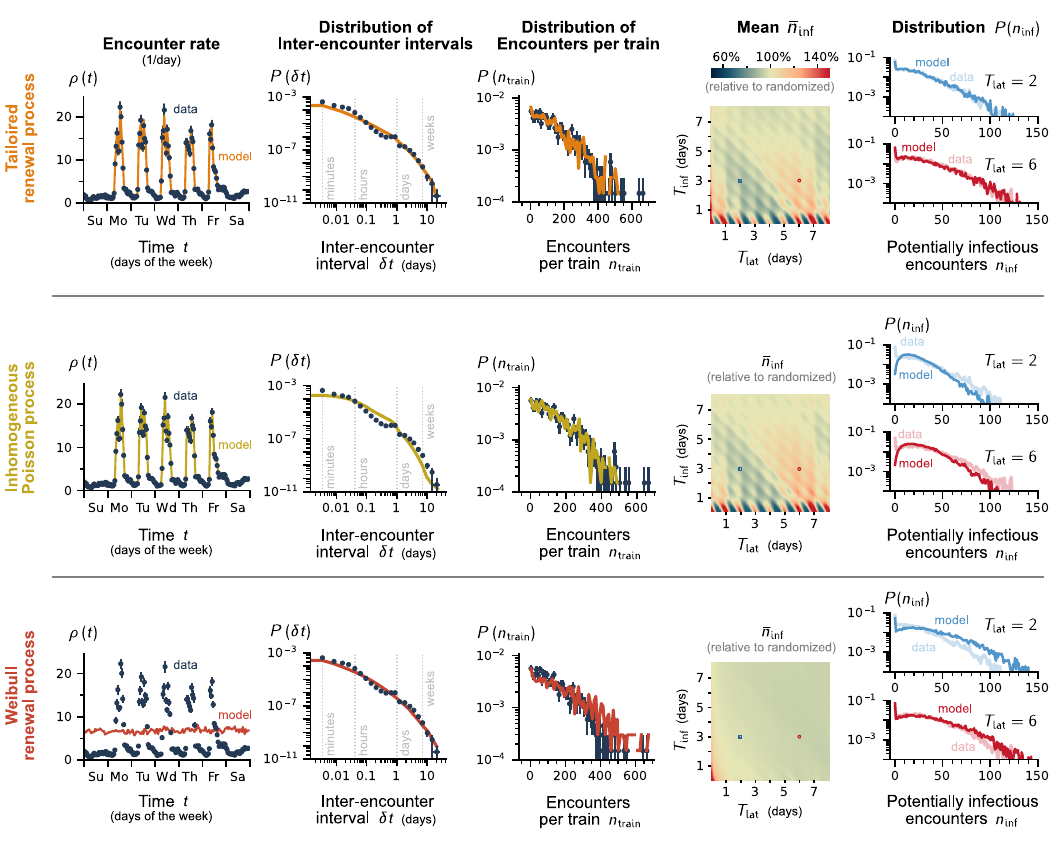}}
    \caption{
    \textbf{Specific temporal statistics of surrogate point processes can be related to specific characteristics of disease spread.}
    \textbf{Top:}~Tailored renewal process that captures time-dependent $\rho(t)$ (first column), heavy-tailed $P(\delta t)$ (second column), and heterogeneous $P(n_\mathrm{train})$ (third column), reproduces core characteristics such as the modulation of $\overline{n}_\ift$ (fourth column, cf. Fig.~\ref{fig3} and Supplementary Material for additional figures for $\Psi(\tau)$) and the increased probability of zero-spreading in $P(n_\ift)$ (fifth column, cf. Fig.~\ref{fig2}).
    \textbf{Center:}~An inhomogeneous Poisson process with only time-varying, cyclostationary $\rho(t)$ and heterogeneous $P(n_\mathrm{train})$ reproduces the modulations in $\overline{n}_\ift$, but underestimates the probability of low $n_\ift$ due to a lack of clustering and the corresponding large inter-cluster times in $P(\delta t)$.
    \textbf{Bottom:}~A Weibull-renewal process with contact clustering from $P(\delta t)$ and heterogeneous $P(n_\mathrm{train})$ reproduces the high probability of rare outcomes in $P(n_\ift)$, but cannot reproduce the modulations in $\overline{n}_\ift$ due to a lack of cyclic temporal structure.
    If, additionally, one relaxes the constraint of heterogeneous $P(n_\mathrm{train})$ and considers statistically identical trains, then both surrogate processes underestimate the probability of large $n_\ift$ related to super-spreading (see Supplementary Material).
    }
    \label{fig4}
\end{figure*}

In the following, we uncover the origin of these periodic modulations using what we call \textit{conditional encounter rate} $\Psi(\tau)$.
In short, $\Psi(\tau)$ describes the average rate of encounters conditioned on an initial encounter (Fig.~\ref{fig3}c).
Considering the initial encounter as an infection, $\Psi(\tau)$ measures the rate of potentially infectious encounters but neglects variability and dispersion.
We find that $\Psi(\tau)$ features a peak at \SI{0} (which implies strong clustering~\cite{barabasi_origin_2005}) and the anticipated periodic modulations between high and low encounter rates (which cause a time-dependent secondary attack rate).
Both, the initial peak and modulations are again lost for randomized controls (yellow line).

Note that we can directly obtain an estimate of $\overline{n}_\ift$ for a particular disease progression ($T_\lat$, $T_\ift$) by integrating $\Psi(\tau)$ over the infectious period (Fig.~\ref{fig3}c, shaded areas).
Reconsidering the previous examples, this explains the lower $\overline{n}_\ift$ for $T_\lat = \SI{2}{days}$ (blue area covering the valley) and the larger $\overline{n}_\ift$ for $T_\lat = \SI{6}{days}$ (red area covering the 7-day peak).
The examples showcase that $\overline{n}_\ift$ is determined by the alignment of the infectious period with regions of low or high $\Psi(\tau)$.

Consequently, since $\overline{n}_\ift$ is related to $R_0$, the interplay between contact patterns and disease progression modulates the pace of epidemic spread.
To illustrate this, we construct a continuous-time branching process, where each exposed individual generates encounters according to $\Psi(\tau)$.
During the infectious period, encounters again have a probability $p_\ift$ to become infected (Fig.~\ref{fig3}d).
Assuming an outbreak that survived the initial generations, we prepare the system with $100$ random initial infections in the interval  $[-T_\lat-T_\ift,0]$.
The resulting time evolution of daily new cases shows clear exponential growth, where the growth rate $\lambda$ trivially decreases with the generation time and, thus, $T_\lat$ (Fig.~\ref{fig3}e).
However, this expected decrease of $\lambda$ for memoryless encounter timings (yellow) is modulated in the model due to modulations in $\Psi(\tau)$, which results in slower-than-random (blue) or faster-than-random (red) growth.

Summarizing, human contact patterns cause a dependence of $\overline{n}_\ift$ on $T_\lat$ that modulates $R_0$ and thereby the growth rate of an epidemic outbreak.

\subsection*{Destabilization and modulation of epidemic spread can be attributed to specific temporal statistics of contact patterns}


After illustrating that non-Markovian statistics can destabilize and modulate epidemic outbreaks, it seems natural to ask how they can be included in models of disease spread.
In such models it is common to approximate encounter times between individuals as memoryless (Poisson) processes~\cite{pastor-satorras_epidemic_2015}.
Assuming independence, these processes can be merged to result in encounter trains with Poisson statistics --- the same statistics as our randomized encounter trains.
In the following, we construct encounter trains with non-Markovian statistics and identify three specific features of contact patterns that are necessary to reproduce the relevant statistics of encounters.
As a proof of principle, we showcase a novel \textit{inhomogeneous Weibull-renewal process} that is constrained by data and reproduces all salient features (Fig.~\ref{fig4}, top row):

\textbf{i)}~Focusing on temporal statistics, the encounter rate $\rho(t)$ averaged across individuals and weeks is time-dependent but cyclostationary;
$\rho(t)$ repeats in a weekly cycle with differences between day and night, and between weekdays and weekends (Fig.~\ref{fig4}, first column).
This can be captured by an inhomogeneous Poisson process (Fig.~\ref{fig4}, middle row), which reproduces the periodic modulation of $\overline{n}_\ift$ (fourth column) and $\Psi(\tau)$ (see Supplemental Fig.~\ref{fig:sm_features_explained}).

\textbf{ii)}~The distribution of inter-encounter intervals $P(
\delta t)$ has high probability for small $
\delta t$ and a heavy tail of
non-vanishing probability for large $\delta t$ (second column).
Because this tail corresponds to long periods without any encounter, it causes the high probability of $n_\ift \approx 0$ (last column) that strongly contributes to the destabilization of epidemic outbreaks.
$P(\delta t)$ is dominantly shaped by the clustering of human contacts and can be well approximated by a Weibull distribution~\cite{van_mieghem_non-markovian_2013, jiang_calling_2013}.
Accordingly, a Weibull-renewal processes (last row) well reproduces $P(\delta t)$ and $P(n_\ift)$,
but it does not have a time-varying $\rho(t)$ and cannot reproduce the period modulations of $\Psi(\tau)$ and $\overline{n}_\ift$.

\textbf{iii)}~Encounter rates vary between individuals (third column).
This variability can be attributed to intrinsic differences in contact behavior (cf.~Fig.~\ref{fig2}, gray vs yellow) and is partly captured by the degree distribution of the contact network~\cite{newman_networks_2010}.
Recall that such across-individual variability is crucial to reproduce the heavy tails of $P(n_\ift)$ and offspring distributions (see also Supplemental Fig.~\ref{fig:sm_features_explained} for generative processes where individuals share a common rate).

%
Clearly, models of disease spread can benefit from a generative process that reproduces those relevant features of human contact patterns, such as the tailored Weibull-renewal process showcased here.
However, although our process reproduces all discussed features, it is built heuristically, and future work is needed to construct microscopic models that give rise to cyclostationary rates with clustering in a principled way, while remaining mathematically tractable.




\section{Discussion}

We analyzed real-world human contact patterns and found that their non-Markovian timings shape epidemic spread in two important ways.
Firstly, they increase the over-dispersion of offspring distributions, compared to random (Poisson) contact patterns, which (i) leads to more zero- and super-spreading events, and (ii) decreases the probability of an epidemic outbreak from an initial infection.
While clustering is typically associated with super-spreading events, it inevitably causes periods of low contact rate that increase the probability of zero-spreading events.
The resulting increase in extinction probability (despite super-spreading) is consistent with previous results, where individual variation of $R_0$ captured the over-dispersion of empirical offspring distributions \cite{lloyd-smith_superspreading_2005}.
Still, the sources of this variation remained poorly understood, with candidates ranging from environmental factors (behavior, seasonality) to intrinsic ones (viral load, susceptibility)~\cite{chen_understanding_2021}.
Here, we disentangled two sources based on contact patterns and identified heterogeneous contact rates and the non-Markovian timing of contacts as relevant factors for over-dispersion in disease transmissibility.

Secondly, human contact patterns non-trivially modulate the pace of epidemic spread depending on the latent period,
which we attribute to time-dependent but cyclostationary encounter rates.
A cyclostationary rate leads to periods of statistically high and low encounter rates conditioned on a potential infection.
How these periods align with the infectious period is affected by the latent period and determines whether the number of potential secondary infections, and in turn $R_0$, increases or decreases.
This modulation of $R_0$ can thus be understood as a resonance following either a constructive or destructive interference between a periodically changing contact rate and the disease progression.
This resonance is a new mechanism to explain the previously observed slow-down or speed-up of diffusion processes on temporal networks due to non-Markovian characteristics~\cite{scholtes_causality-driven_2014}.

In the main manuscript we focus on deterministic disease progression with fixed periods ($T_\lat$, $T_\ift$), but we also considered non-deterministic disease progression with gamma-distributed
periods~\cite{bailey_stochastic_1964, anderson_spread_1980,
lloyd_destabilization_2001}; the results are summarized in the Supplementary Material (Fig.~S1).
We find our main conclusion  verified for non-deterministic disease progression:
the probability of zero-spreading events is reliably higher for human contact patterns compared to randomized;
however, the modulation of $\overline{n}_\ift$ with $T_\lat$ is smeared out with increasing variability in the period durations.
Thus, in the unrealistic (but commonly adopted) limit of exponentially distributed periods, human contact patterns still reduce the robustness of outbreaks but no longer modulate the pace of epidemic spread.

To reproduce the relevant temporal features of human contact patterns, we introduced non-Markovian contact dynamics in the form of Weibull-distributed inter-encounter intervals (clustering) or inhomogeneous encounter rates (cyclostationarity).
Previous studies of non-Markovian disease spread~\cite{van_mieghem_non-markovian_2013} found that clustering drastically affects the epidemic threshold for $T_\lat=0$, which is caused by the high frequency of small inter-encounter intervals~\cite{masuda_small_2020} that, in our context, manifests as a near-zero peak in the conditional encounter rate (Fig.~\ref{fig3}).
Although it was shown that some non-Markovian models can be mapped onto effective Markov models~\cite{starnini_equivalence_2017,feng_equivalence_2019}, our results suggest that the non-Markovian and cyclostationary features of human contact patterns make a similar general mapping elusive.
This highlights the necessity for generative models that are non-Markovian, yet well understood and simple enough to find broad use in epidemic modeling and beyond.

Our work is a first step towards providing such models.
We identified temporal statistics of real-world contact data that affect disease spread, and faithfully reproduced them with our tailored Weibull-renewal process.
Thereby, our work provides an accessible pathway towards including non-Markovian statistics into spreading processes, in general, and paves the way to systematically study their non-equilibrium physics.

\section*{Acknowledgements}
We would like to thank Peter Sollich and Sune Lehmann for helpful discussion.
J.Z.~received financial support from the Joachim Herz Stiftung. J.Z., F.P.S., J.D., V.P., and M.Wi.~acknowledge funding by the Max Planck Society. F.P.S. and V.P. acknowledge funding from the SFB 1528 \enquote{Cognition of Interaction}.

\newcommand{\method}[1]{\textbf{#1}}

\newpage
\section*{Methods}
\method{Extracting contacts from real-world physical proximity data:}
Consider data composed of a list of co-locations (physical proximity) described by the tuple (timestamp, user id A, user id B).
We first sort the co-location times into unique lists for all id pairs (A,B) and (B,A).
For each valid A, we then iterate over its list of (A,B) and merge co-location times that span consecutive time steps to construct pairwise contacts with starting time $s$ and duration $D$.
Combining these contacts yields a list of contacts $\{(s_i,D_i)\}_{A}$ for each participant $A$.

From the lists of contacts, we construct a point-process-like representation for each participant that we call \textit{encounter train} (see Fig.~\ref{fig1}).
Throughout the manuscript, an \textit{encounter} refers to the starting time of a contact.
The encounter train of participant A is the time-sorted list of all contact starting times $s_i$ and can formally be written as
\begin{equation}
    T(t) = \sum_i \delta(t-s_i)
\end{equation}

The main data set from the Copenhagen Networks Study~\cite{sekara_fundamental_2016, sapiezynski_interaction_2019} is based on Bluetooth signals between phones of individuals that participated in the study.
The published data is a list of interactions described by the tuple (timestamp, A, B, RSSI), where user id B can be negative if the interaction is with a device outside of the study or an empty scan, and RSSI is the received (Bluetooth) signal strength indicator.
The RSSI can be considered as a proxy for interaction distance, especially since all participants used the same device~\cite{sekara_strength_2014}, with an $\text{RSSI}\approx \SI{-80}{dBm}$ corresponding to a distance of about $\SI{2}{m}\pm\SI{0.5}{m}$.
Since the data provides a maximal $\text{RSSI}$ per time window, we consider $\text{RSSI}<\SI{-80}{dBm}$ to indicate interactions to be further apart than $\SI{2}{m}$ throughout the full time window~\cite{sekara_strength_2014}, and exclude them.
Consequently, we filter the raw data to only include those interactions that are within the study (user id B $\geq 0$) and have $\text{RSSI}\geq\SI{-80}{dBm}$.
The data set covers a duration of $t_\mathrm{max}=\SI{28}{days}$, with a time step of $\SI{5}{min}$, for 675 encounter trains.


\method{Average time-dependent encounter rate $\rho(t)$:}
Because encounter trains are a point-process-like representation, we can define an encounter rate as the number of encounters in a window of size $\Delta t$.
Assuming statistical independence between weeks and between participants, we determine the average time-dependent encounter rate $\rho(t)$ by averaging the number of encounters in a time windows of size $\Delta t=\SI{1}{h}$ throughout the week (i.e.~first hour of a Sunday until last hour of a Saturday) across weeks of the experiment and across participants.
Statistical errors are calculated on the level of participants using delete-$m$ jackknife error analysis.


\method{Inter-encounter interval $\delta t$:}
To study temporal clustering and contact bursts, we measure the interval $\delta t$ between consecutive encounter times.
Since we are interested in the encounter statistics, each encounter has the same statistical weight independent of its encounter train origin.
Consequently, the distribution $P(\delta t)$ is simply the distribution over all observed intervals.
To estimate statistical errors, we take into account that the number of encounters $n_j$ differs between individual trains (hence also the number of inter-encounter intervals $n_j-1$), and evaluate statistical errors on the level of observed intervals using delete-$m_j$ jackknife error analysis with $m_j=n_j-1$.

\method{Conditional encounter rate $\Psi(\tau)$:}
To investigate how contact patterns interact with disease spread, we measure the encounter rate $\Psi(\tau)$ upon a hypothetical infection from an encounter at $\tau=0$.
To construct $\Psi(\tau)$, we iterate over all encounters to measure the time-dependent encounter rate with temporal resolution of the experiment, starting from the encounter time, i.e, $\tau=t-s_i=0$, until $\tau=\tau_\mathrm{max}$ (we typically chose $\tau_\mathrm{max}=\SI{10}{days}$) or, if $t_\mathrm{max}-s_i < \tau_\mathrm{max}$, until $\tau=t_\mathrm{max}-s_i$.
We then average over all these time-dependent encounter rates taking into account their different lengths.
To estimate statistical errors, we take into account that the number of encounters $n_j$ differs between individual trains by using delete-$m_j$ jackknife error analysis with $m_j=n_j$.

\method{Disease model:}
We consider a disease that progresses in three discrete states upon infection: exposed-infectious-recovered.
The duration $T_\lat$ within the exposed state is called \textit{latent period} and the duration $T_\ift$ within the infectious state is called \textit{infectious period}.
For our main results, we consider the simple and intuitive case of a \textit{deterministic disease progression}, where these periods are always the same.
This corresponds to drawing the periods from delta distributions, which is quite different to commonly employed approximations that draw periods from exponential distributions (as expected for Poisson processes that describe many state transitions, from radioactive decay to chemical reactions).
To control that our results also apply to \textit{non-deterministic disease progression}, we repeated our analysis for the more realistic case of gamma-distributed periods~\cite{bailey_stochastic_1964, anderson_spread_1980, lloyd_destabilization_2001} and obtained consistent results (Supplemental Material).


\method{Potentially infectious encounters $n_\ift$:}
To avoid assumptions on the probability of infection upon encounter, we introduce potentially infectious encounters as the number of encounters that occur during the infectious period of a hypothetical disease.
For the deterministic disease progression, we can enumerate the statistics by iterating over all encounters of the data set.
For each encounter $s_i$, we check whether the disease progression still fits into the experimental duration $t_\mathrm{max}$ ($s_i+ T_{\rm lat} + T_{\rm inf} \leq t_\mathrm{max}$), and if true, estimate $n_\ift$ as the number of subsequent encounter $s_j$ for which $T_{\rm lat} < s_j - s_i < T_{\rm lat}+T_{\rm inf}$.
For the non-deterministic disease progression, we need to sample disease realizations (see Supplemental Material).
Statistical errors are calculated again on the level of encounters using the delete-$m_j$ jackknife analysis with $m_j=n_j$.

\textbf{Branching process with empirical distribution:}
To estimate the survival probability from the empirical distribution of potentially infectious encounters, $P(n_\ift)$, we construct a discrete-time data-driven branching process (Fig.~\ref{fig2}c).
In a first step, each infection causes $X\sim P(\potInfEnc)$ potentially infectious encounters.
In a second step, each of these encounters can cause a secondary infection with probability $p_\ift$, such that the number of secondary infections is binomial, $Y\sim\mathcal{B}(X,p_\ift)$.
From $Z_{t}$ infections in generation $t$, we thus obtain $Z_{t=1}=\sum_{i=1}^{Z_{t}} Y_i$ infections in the next generation.

\textbf{Continuous-time branching process with inhomogeneous contacts:}
To study the pace of epidemic spread, we construct a continuous-time branching process that captures the conditional encounter rates but neglects interactions between infected.
Here, each infected individual generates independent encounter trains starting from the initial infection time as inhomogeneous Poisson processes with a time-dependent rate given by the conditional encounter rate (Fig.~\ref{fig3}d).
Only those encounters that occur during the infectious period cause secondary infections with a chosen probability $p_\ift$.
Every secondary infection then generates a new encounter train and so on.
For simplicity, we restrict our example to deterministic diseases with fixed latent and infectious periods.

\method{Point process models to approximate human contact patterns:}
To disentangle the effect of distinct features of human contact patterns on the statistics of encounters, we constructed point-process models that captured (i) the distribution of rates across individuals, (ii), a time-dependent average encounter rate, and (iii), the distribution of inter-encounter intervals, or a combination thereof (see Supplementary Material for comparison of combinations).

To reproduce the inter-individual variability, we consider the same number of encounter trains as present in the data and weight each train with their relative rates, i.e, $w_i=n_{\mathrm{train},i}/\langle n_\mathrm{train}\rangle$, where $\langle \cdot\rangle$ is the average across trains.

To reproduce a time-dependent encounter rate $\rho(t)$, we employ thinning~\cite{lewis_simulation_1979}: From a hidden process with rate $\rho_i=\max_t[\rho(t)]$ we accept events at time $t$ with probability $p_i(t) = \rho(t)/\rho_i$.
This procedure can formally only be applied for memory-less hidden processes, i.e., Poisson processes, in which case it results in an \textit{inhomogeneous Poisson process}.
To further reproduce heterogeneous rates in the inhomogeneous Poisson process, we fix $p(t)$ but rescale the rates of the hidden processes, $\rho_i = w_i \max_t[\rho(t)]$.

To reproduce the empirical distribution of inter-encounter intervals, we construct a \textit{Weibull-renewal process}:
Inter-encounter intervals are drawn from a Weibull distribution with scale parameter $\lambda$ and shape parameter $k$.
The Weibull distribution was parameterized by a fit to the data yielding $(k,\lambda)=(0.3690,3030)$.
To further reproduce heterogeneous rates in the Weibull-renewal process, we notice that the mean rate of a Weibull-renewal process is given by $\rho_i=\left[\lambda_i\Gamma(1+1/k_i)\right]^{-1}$, such that we can simply choose $k_i=k$ and $\lambda_i=\lambda/w_i$.

To combine all features in a single model, we construct a \textit{tailored renewal process}: Weibull-renewal processes with heterogeneous rates and additional (heuristic) thinning.
We start with a set of hidden Weibull-renewal processes with $k_i=k$, $\lambda_i=\lambda/w_i$, and time-dependent acceptance probability $p(t)$ with time-average $\overline{p(t)}$.
The mean rate of each process is $\rho_i = \overline{p(t)}w_i/\lambda\Gamma(1+1/k)$.
Since we cannot fit $(k,\lambda)$ of the hidden process, we further constrain the parameters with the mean rate from data, i.e., $\overline{\rho(t)} = \langle \rho_i\rangle = \overline{p(t)}/\lambda\Gamma(1+1/k)$, with $\langle w\rangle = 1$ by construction.
Since $\overline{\rho(t)}/\overline{p(t)}=\max_t[\rho(t)]$, we thus find $\lambda = \left[\max_t[\rho(t)]\Gamma(1+1/k)\right]^{-1}$, such that $k$ remains the only free parameter.
We obtained our best estimate of $k$ by minimizing the Kullback-Leibler divergence~\cite{kullback_information_1951} between the distribution tails ($\delta t \gtrsim \SI{0.5}{days}$) of model and empirical $P(\delta t)$, finding $k\approx0.24$.

\method{Jackknife error estimation:}
To estimate statistical errors of our results, we use jackknife error estimation while carefully taking into account the size of the left-out data set.
The basic idea of the jackknife method is to estimate from some data $X=\{x_1,\ldots,x_g\}$ the variance of an observable $\hat{O}=f(X)$ using a systematic resampling approach~\cite{efron_jackknife_1982}.
Jackknife samples $O_j$ are generated by systematically leaving out data, e.g., $\hat{O}_j=f(X_{\bar{\jmath}})$ with $X_{\bar{\jmath}} = \{x_1,....,x_{j-1},x_{j+1},...x_g\}$.
Importantly, here each $x_j$ can be a block of (differently many) data points.
While typically theses blocks have the same size $m$ (delete-$m$ jacknife), they could have different sizes $m_j$ (delete-$m_j$ jackknife), which will be relevant for some of our cases.
From the jackknife samples, one can show that bias-reduced estimators of the mean and variance are given by~\cite{busing_delete-m_1999}

\begin{align}
    \hat{O}_J &= \sum_{j=1}^g \frac{1}{h_j} \left(h_j \hat{O} - (h_j-1) \hat{O}_j\right),\nonumber\\
    \hat{\sigma}^2_J &= \frac{1}{g}\sum_{j=1}^g \frac{1}{h_j-1}\left(h_j \hat{O} - (h_j-1) \hat{O}_j - \hat{O}_J \right)^2\label{eq:delete_mj},
\end{align}
where $h_j=(\sum_{j=1}^g m_j)/m_j$, and $\hat{O}=f(X)$ is the naive estimate.
For blocks of equal size, $m_j=m$, we have $h_j=g$ and this simplifies to

\begin{align}
    \hat{O}_J &= g \hat{O} - \frac{g-1}{g}\sum_{j=1}^{g} \hat{O}_j,\nonumber\\
    \hat{\sigma}^2_J &= \frac{g-1}{g}\sum_{j=1}^g\left(\hat{O}_j - \frac{1}{g}\sum_{j=1}^g \hat{O}_j\right)^2\label{eq:delete_m}.
\end{align}
In our case, the data $X$ is the set of all encounter trains and in the resampling step we leave out individual encounter trains.
Since trains include differently many encounters, this can results in removing blocks of different sizes.
In particular, all observables that derive from the number of encounters, e.g., $\overline{n}_\text{inf}$ or $P(n_\text{inf})$, require the delete-$m_j$ analysis, Eq.~\eqref{eq:delete_mj}, to estimate the statistical error.
On the other hand, for observables that depend on time-binned data, e.g., the time-dependent rate, each encounter train has the same size given by the number of time bins during the recording such that the delete-$m$ analysis, Eq.~\eqref{eq:delete_m}, is sufficient to estimate the statistical error.

\method{Data availability:}
Our methods are publicly available ~\cite{noauthor_resonance_contact_disease_2021} and applied to open-access data-sets~\cite{sapiezynski_interaction_2019, genois_can_2018}.

\bibliography{ContactBehaviorDisease}

\clearpage
\renewcommand{\thefigure}{S\arabic{figure}}
\setcounter{figure}{0}
\renewcommand{\thetable}{S\arabic{table}}
\setcounter{table}{0}

\onecolumngrid
\section*{Supplementary Material}
In the Supplementary Material, we provide additional controls to verify the robustness of our results in the main manuscript.

\subsection{Non-deterministic disease progression}
\label{sec:gamma_disease}

\begin{figure}[b!]
    \centerline{\includegraphics[]{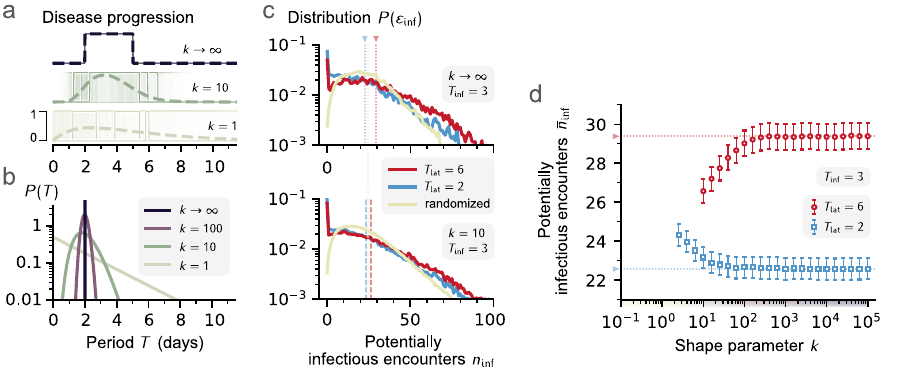}}
    \caption{\textbf{Increased variability of disease-stage periods does not affect conclusions on robustness of outbreaks but weakens the modulation of spreading pace.}
    \textbf{a,~b:}~To include person-to-person variability, we draw both latent and infectious periods from gamma distributions characterized by the shape parameter $k$, which interpolates between exponential ($k=1$) and delta ($k\to\infty$) distributions.
    For small $k$, $T_{\rm lat}$ and $T_{\rm inf}$ differ in duration from realization to realization --- across individuals, the probability to be infectious at a given time is smeared out.
    As $k$ increases, the periods vary less around their expected value and the disease progression eventually becomes deterministic.
    \textbf{c:}~Probability distributions of potentially infectious encounters $n_{\rm inf}$
    for $k \to \infty$ (top) and $k=10$ (bottom).
    When randomizing trains, the probability of zero-infectious encounters is suppressed.
    \textbf{d:}
    Mean number of potentially infectious encounters $\overline{n}_{\rm inf}$ as a function of $k$ for the two examples from the main manuscript ($T_\ift = \SI{3}{days}$ and $T_\lat = 2$ or $\SI{6}{days}$).
    For $k\to\infty$, we recover the result for deterministic disease progression (dashed lines), where the latent period induces a notable difference between the two examples.
    For smaller $k$, the resonance effects remain relevant but the difference decreases, see also Fig.~\ref{fig:sm_dispersion_fast_progression}.
    }
    \label{fig:sm_dispersion}
\end{figure}

In the main manuscript, we have focused on deterministic disease progression, where latent and infectious periods had a precise duration.
For a more realistic view, we want to allow for the latent and infectious periods to vary from case to case (Fig.~\ref{fig:sm_dispersion}).
To that end, we draw the periods $T_{\rm lat}$ and $T_{\rm inf}$ from a gamma distribution, but we keep the mean duration fixed\footnote{%
Since now gamma-distributed periods are stochastic, we can no longer enumerate the statistics of $P(n_\ift)$, but have to sample it.
Specifically, we sample $10^6$ disease realizations, where we first draw a random realization of the disease progression $(T_{{\rm lat},i}, T_{{\rm inf},i})$, to then draw disease start times $s_i$ from the ensemble of all encounters until we find an $s_i$ such that the disease progression is within the remaining duration of the experiment,  i.e., $s_i + T_{{\rm lat},i} + T_{{\rm inf},i} \leq T_{\rm exp}$.
Only once we have a valid disease start time $s_i$, we count the number of subsequent encounters within the infectious period as above.
If, for any disease realization, we need to draw more than 1000 disease start times until we find a valid one, we abort the estimation for the set of parameters $(T_{\rm lat},T_{\rm inf},k)$.
By this procedure to first draw and fix a random realization of the disease progression, we avoid a bias towards small periods that would occur due to the finite period of the experimental data.
}.
The case-to-case variability is then parameterized through the shape parameter $k$ (Fig.~\ref{fig:sm_dispersion}a).

In particular, we fix the mean $\langle T_i\rangle$ by choosing the scale $\theta = \langle T_i\rangle/k$ such that
$P_k(T_i) = T_i^{k-1}e^{-T_i/\theta}/[\Gamma(k)\theta^k]$.
The shape parameter allows us to interpolate between a delta distribution ($k\to\infty$) and an exponential distribution ($k=1$), as commonly assumed in computational epidemiology for mathematical tractability~\cite{pastor-satorras_epidemic_2015}.

Clinically observed distributions of periods between disease states are neither delta distributed nor exponential distributed and may be best described by distributions with a clear peak but vanishing probability at zero, such as log-normal distributions or gamma distributions with shape parameters in between $(1,\infty)$.
On the one hand, delta-distributed periods seem like a convenient but unrealistic simplification.
On the other hand, exponentially distributed periods may appear more realistic, however, they imply an artificially high probability of short durations, which in turn leads to realizations where the infectious period either starts shortly upon infection or has close-to-zero duration (example traces in Fig.~\ref{fig:sm_dispersion}a).
In fact, it has been argued already in the past that gamma distributions are more realistic~\cite{bailey_stochastic_1964, anderson_spread_1980, lloyd_destabilization_2001}, see for example empirical distributions of latent periods for COVID-19~\cite{xin_estimating_2022}, such that more realistic shape parameters could be in the range $h\in[5,20]$ which is between delta and exponential.
To investigate how case-to-case variability affects the number of potentially infectious encounters $n_{\rm inf}$,
we again consider the probability distribution $P(n_{\rm inf})$ (Fig.~\ref{fig:sm_dispersion}c), and revisit the two examples
($T_{\rm inf} = \SI{3}{days}$ with either $T_{\rm lat} = \SI{2}{days}$ or $T_{\rm lat} = \SI{6}{days}$).

For the delta-disease ($k\to\infty$, top), the red ($T_{\rm lat} = \SI{6}{days}$) and blue ($T_{\rm lat} = \SI{2}{days}$) distributions exhibit a peak at zero, they are broad, and have a long tail.
Comparing red and blue, we find that the latent period determines the height of the peak at zero as well as the position of the bulk distribution, and, thereby, determines the mean number of potentially infectious encounters (dashed vertical lines).
The mean values clearly differ.
Again comparing with the randomized encounter trains (yellow line), we find no peak at zero, a shorter tail, and no dependence on the latent period (the respective randomized lines fall on top of each other).
Importantly, a peak at zero implies that the infected individual does not pass on the infection, so that the disease becomes more likely to die out if case numbers are low.

Changing to the non-deterministic disease progression ($k=10$, Fig.~\ref{fig:sm_dispersion}c bottom),
the distribution from randomized data is barely affected.
However, the red and blue distributions are more similar to each other, but also broader and smoother than for the delta-disease.
This can be explained by gamma-distributed periods acting as a smoothing kernel along both dimensions of Fig.~\ref{fig:sm_dispersion}f, where variability in the infectious period directly affects $\overline{n}_\text{inf}$, while variability in the latent period affects $\overline{n}_\text{inf}$ through the resonance effect.
Consequently, we expect that with decreasing shape parameter $k$, the mean number of potentially infectious encounters becomes independent of the mean latent period.
Indeed, when considering $\overline{n}_\text{inf}$ as a function of $k$, we find that the estimates for our two examples approach each other, as $k$ decreases (Fig.~\ref{fig:sm_dispersion}d).

Note that our analysis has a lower bound in $k$ once realizations of disease progression (latent + infectious period) cannot find sufficiently many initial encounters to fit into the finite duration of the experiment (4 weeks for CNS).
However, using other examples with smaller latent and infectious periods (where we can acquire enough statistics), we show that the two extreme cases meet for $k\approx 1$ (see Fig.~\ref{fig:sm_dispersion_fast_progression}).

\subsection{Non-deterministic disease progression with fast disease stages}
\label{sec:complement_fast}

With the example of the main text, we were not able to sample the for small $k$ values (more variability across disease realizations) because the 28-day duration of the data becomes too short once the periods of disease progression are close-to exponentially distributed ($k \to 1$).

To avoid this issue and to illustrate shorter timescale, we here compare with another hypothetical example of a ``faster'' disease progression, where $T_\text{inf} = \SI{0.5}{days}$ and $T_\text{lat}$ is either $1$ or $1.5$ days (Fig.~\ref{fig:sm_dispersion_fast_progression}).
In this case, expected periods are very short so that even realizations with periods that severely exceed their expected value fit into the 28-day duration.

We find a few noteworthy aspects:
First, the absolute value of potentially infectious encounters $n_\mathrm{inf}$ is much lower for faster disease progression.
This is due to the much shorter infectious period.
However, the relative deviation from the randomized surrogate data is consistent between the two examples.
Second, we now find that the longer latent period ($1.5$ vs. $1.0$) leads to fewer potentially infectious encounters (red vs.~blue).
This is just a result of the chosen latent periods: choosing an even longer latent period would result in a respective decrease of $n_\mathrm{inf}$ (e.g. $2.0$ vs $1.5$).
Third, as expected, as $k \to 1$, the estimates of $n_\mathrm{inf}$ overlap for different latent periods, because individual disease realizations become very variable (cf.~Fig.~\ref{fig:sm_dispersion}a).

\clearpage
\begin{figure}[h]
    \centerline{\includegraphics[]{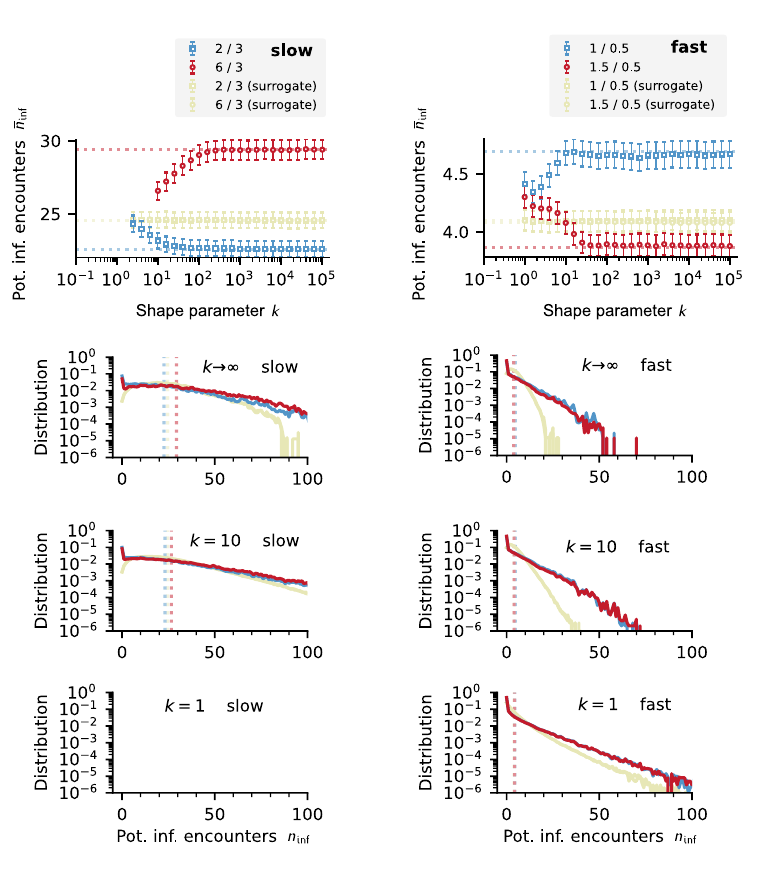}}
    \caption{\textbf{Example with smaller latent and infectious periods provides additional insight on low-$k$ regime of non-deterministic disease progression}. Here, we compare the example from the main manuscript (\enquote{slow}, $T_\text{inf} = \SI{3}{days}$ and $T_\text{lat}$ either $2$ or $6$ days) with a hypothetical \enquote{fast} disease progression ($T_\text{inf} = \SI{0.5}{days}$ and $T_\text{lat}$ either $1$ or $1.5$ days) and show both the mean number of potentially infectious encounters as a function of the shape parameter $k$ of the gamma-distributed latent and infectious period (top) as well as their distributions for selected $k$.
    Due to the finite duration of the recording, the accessible low-$k$ regime is determined by the mean latent and infectious period, because for low $k$ large periods quickly exceed the finite duration.
    For faster disease progression (smaller latent and infectious period), we observe modulations on smaller timescales and, in addition, reach the low-$k$ regime of exponentially distributed periods ($k=1$) commonly assumed in epidemiological simulations.
    As one can see more clearly for faster disease progression, the mean number of potentially infectious encounters approach each other in this low-$k$ regime, which can only be anticipated from the results for slower disease progression.
    This implies that modulations will not be present for exponentially distributed latent and infectious periods but only for more realistic non-exponentially distributed ones with higher $k$.
    }
    \label{fig:sm_dispersion_fast_progression}
\end{figure}

\clearpage
\subsection{Simple point processes cannot fully reproduce temporal features}

As we noted in the main text, each of the considered simple generative models is insufficient to reproduce the full set of observed temporal features of contact patterns.
Here, we provide a more complete overview of processes and the features they reproduce (Fig.~\ref{fig:sm_features_explained}).
We identified three relevant features:

    \textbf{i)}~a time-varying, cyclostationary encounter rate $\rho(t)$ (first column)

    \textbf{ii)}~a heavy-tailed inter-encounter-interval distribution $P(\delta t)$ (second column)

    \textbf{iii)}~encounter rates vary between individuals (third column)

Feature i is responsible for modulation of the conditional encounter rate $\Psi(\tau)$ (fourth column).
Combined with disease progression, this causes modulations of $\overline{n}_\ift$ and distributions $P(n_\ift)$ that vary with $T_\lat$ and $T_\ift$ (last two columns).
Feature ii is responsible for zero-spreading events, which destabilize epidemic outbreaks and manifest in the peak of $P(n_\ift)$ for $n_\ift \approx 0$ (last column).
Feature iii is responsible for super-spreading events, which correspond to the tail of $P(n_\ift)$ and, thus, can cause a systematic shift of $\overline{n}_\ift$.

\begin{figure}[h]
    \includegraphics[width=1.0\textwidth]{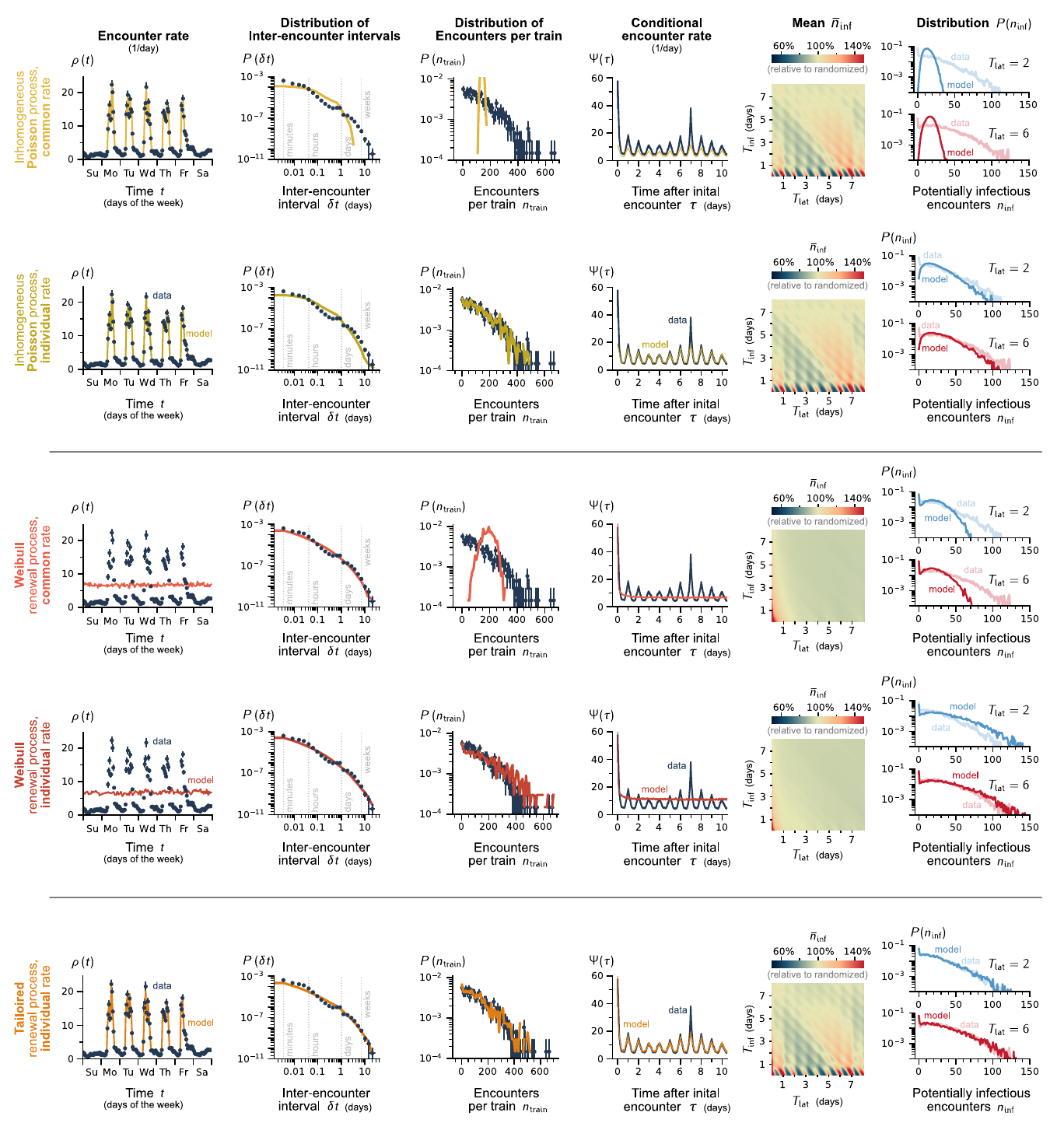}
    \caption{
    \textbf{Simple generative processes are not capable of reproducing all aspects of the full real-world contact statistics.}
    An inhomogeneous Poisson process can reproduce cyclostationary $\rho(t)$ but fails to reproduce sufficiently heavy-tailed $P(\delta t)$.
    A Weibull-renewal process reproduces heavy-tailed $P(\delta t)$ but fails to reproduce the cyclostationary $\rho(t)$.
    Generative processes need to reproduce variability between individuals $P(n_\textrm{train}$) to produce long-tailed distributions $P(n_\ift)$.
    In order to reproduce the data in all the considered aspects, we had to design a tailored Weibull-renewal process with homogeneous rates.
    }
    \label{fig:sm_features_explained}
\end{figure}

\clearpage

\subsection{Control regarding continuous contribution of participants}
\label{sec:control_continuous}
In the main manuscript, we use the full published data set of the Copenhagen Networks Study~\cite{sapiezynski_interaction_2019}, covering the physical proximity data of 675 participants. Upon closer inspection, there are periods both at the beginning and the end of the experiment without entries for some of these 675 participants. Since entries also occur for Bluetooth signals with unknown devices, this may indicate irregularities in the contact behavior of some of the participants, e.g., incomplete participation of individuals.

In order to make sure that our results are not affected by such boundary effects, we reanalyzed the data and included only the contact trains of those individuals for which any Bluetooth signal was recorded on both the first and last day of the study (Fig.~\ref{fig:sm_exclude_incomplete}). Technically, we achieved this easily by restricting our analysis to those IDs for which timestamps were recorded within the first day (timestamp $<1\cdot24\cdot60\cdot60\SI{}{s}$) and the last day (timestamp $>27\cdot24\cdot60\cdot60\SI{}{s}$), reducing the data set to 533 contact trains.

This control analysis fully supports our quantitative results from the main manuscript (Fig.~\ref{fig:sm_exclude_incomplete}) such that we can rule out artifacts from boundary effects of incomplete participation.
In particular, we observe a matching weekly structure of the encounter rate (Fig.~\ref{fig:sm_exclude_incomplete}a), a matching distribution of inter-encounter intervals that can be fitted with a Weibull distribution (Fig.~\ref{fig:sm_exclude_incomplete}b), and a matching conditional encounter rate (Fig.~\ref{fig:sm_exclude_incomplete}c).
Consequently, both mean potentially infectious encounters for deterministic disease progression (Fig.~\ref{fig:sm_exclude_incomplete}d and f) as well as for non-deterministic disease progression (Fig.~\ref{fig:sm_exclude_incomplete}e and g) match our main results.

\begin{figure}[h]
    \centerline{\includegraphics[width=\columnwidth]{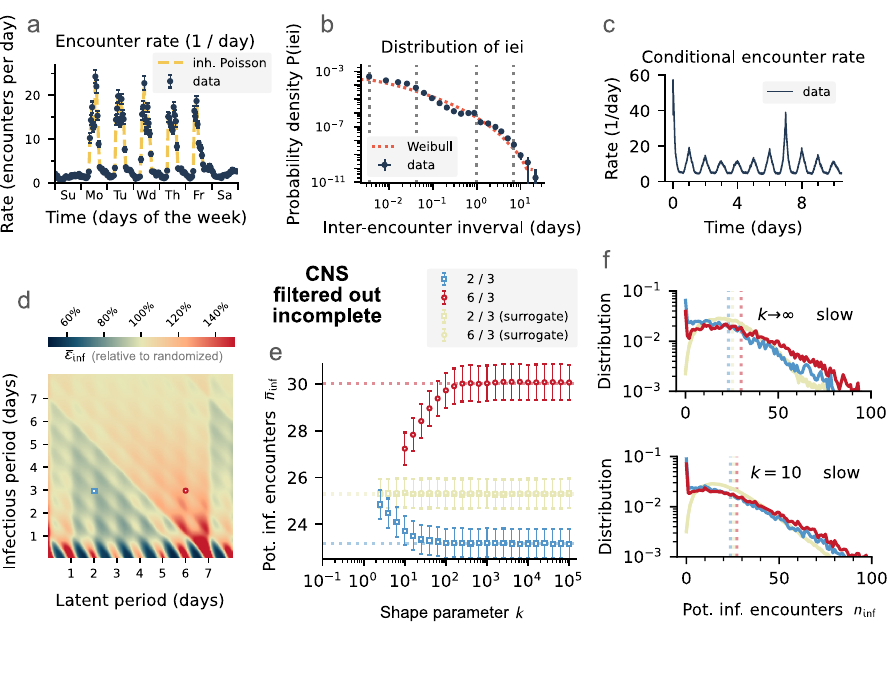}}
    \caption{\textbf{Control regarding continuous contribution of participants.}
    Here, we excluded those trains that did not have any encounters during the first or last day.
    This results in 533 instead of 675 trains.
    Panels match key figures from the main manuscript.
    Results are consistent.
    The fitted Weibull distribution has shape parameter $0.3739$ and scale parameter $3161$.
    }
    \label{fig:sm_exclude_incomplete}
\end{figure}

\clearpage
\subsection{Analysis of an alternative data set}
\label{sec:control_InVS}
To further test the robustness of our results, we repeated our analysis on a completely independent data set. Here, we consider contact data recorded at one of the office buildings of the French Institute for Public Health Surveillance \enquote{InVS}~\cite{genois_data_2015}. This data is recorded with a different technique, namely so-called near-field chips that only record signals in close proximity ($\lesssim\SI{5}{m}$) and avoid to threshold the Bluetooth signal. Moreover, the temporal resolution of contacts is \SI{20}{s} as opposed to \SI{5}{min} in the main manuscript. In addition, the data is recorded for a different social group, namely adults within an office building. Last, the data is recorded in another country (France) by a different collaboration (SocioPatterns).
The data set spans 2 weeks of recording 145 participants (two thirds of the staff agreed to participate).

The analysis of this completely independent data set provides completely consistent results to those presented in the main manuscript (Fig.~\ref{fig:sm_invs}). When comparing the results, we have to highlight that the available statistics for this data set are much smaller due to smaller duration and smaller number of participants.
However, we clearly see the expected weekly structure in the encounter rate (which is here again dominated by working days because of office hours), the distribution of inter-encounter intervals that is well described by a Weibull distribution, as well as the typical conditional encounter rate with a peak at 7 days. Consequently, also the results for (delta) disease progression are consistent with our main findings on the modulation of potential secondary infections (Fig.~\ref{fig:sm_invs}d).
Results on non-deterministic disease progression are confined to shorter latent and infectious periods due to the shorter experimental duration.
We conclude that the additional data set fully supports our results in the main manuscript.

\begin{figure}[h]
    \centerline{\includegraphics[width=\columnwidth]{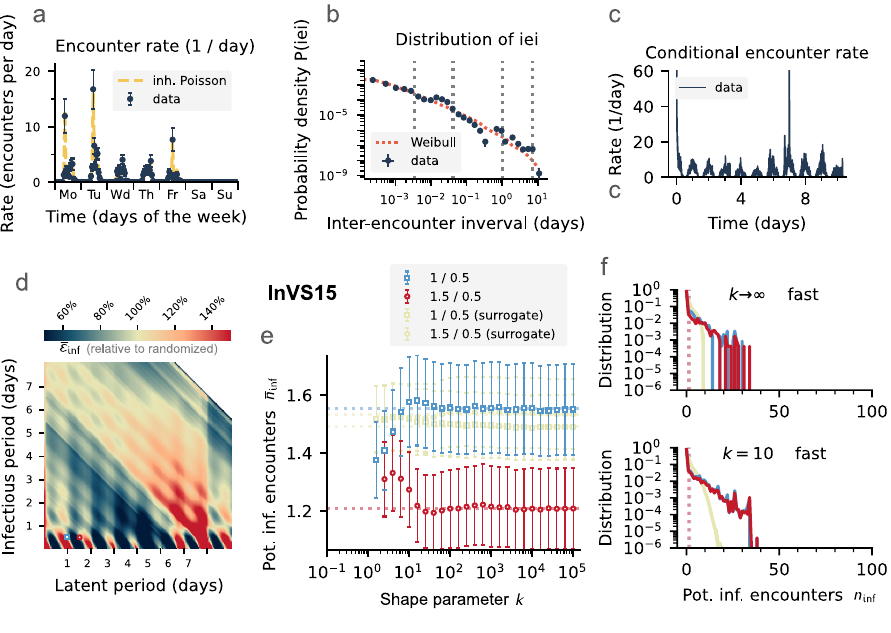}}
    \caption{
    \textbf{Main results for other data set.}
    Because the recording lasted \enquote{only} two weeks, the duration of disease that can be sampled were limited to the fast disease progression.
    The fitted Weibull distribution has shape parameter $0.225$ and scale parameter $675.0$.
    }
    \label{fig:sm_invs}
\end{figure}

\clearpage
\subsection{Infections from outside the study group}
\label{sec:complement_outside}
To check the effect of contacts that could take place with people who were not part of the study,
we investigate disease onsets that do not directly follow the contact patterns observed in the data (Fig.~\ref{fig:controls_infectious_encounters}).
In our analysis of the main manuscript, an infection could only originate from an encounter with another participant in the data set.
Here, we keep the original encounter trains (to evaluate potentially infectious encounters) but the disease onset can occur at any time, due to a hypothetical contact with an external person.
We focus on the resulting distribution and the mean of potentially infectious encounters $n_{\rm inf}$ (Fig.~\ref{fig:controls_infectious_encounters}).

We distinguish the following different possibilities of disease onset:
\begin{description}[style=multiline, leftmargin=3.5cm, labelindent=1cm]
\item[internal] Onsets occur as in the main manuscript only at encounters recorded in the data set. This naturally incorporates the spatio-temporal structure of encounters, in particular their temporal inhomogeneity and their variability across participants.

\item[external i)] Onsets occur completely randomly, at random times for random participants. This neglects both temporal inhomogeneity of encounters and their variability across individuals.

\item[external ii)] Onsets occur at random times with probability proportional to the encounter rate for random participants with probability proportional to their total number of encounters. This incorporates both the (averaged) temporal inhomogeneity of encounters and the variability across individuals.

\item[external iii)] Onsets occur at random times with probability proportional to the encounter rate but uniformly across random participants. This incorporates the (averaged) temporal inhomogeneity of encounters but neglects the variability across individuals.

\item[external iv)] Onsets occur at uniformly random times for random participants with probability proportional their total number of encounters. This neglects the temporal inhomogeneity of encounters but incorporates the variability across individuals.

\end{description}

Once an onset has been chosen, the disease progression is modeled as in the main manuscript.
We focus on the usual examples but use gamma-distributed latent and infectious periods with $k=10$ to evaluate potentially infectious encounters ($k=10$ aims to strike a balance between completely deterministic and exponential disease stages~\cite{bailey_stochastic_1964, anderson_spread_1980, lloyd_destabilization_2001}).

Comparing the different versions of disease onset (Fig.~\ref{fig:controls_infectious_encounters}), we can attribute clear effects to both the temporal inhomogeneity of the onset time as well as the variability of onset times across individuals. Please note that in all cases the encounter statistics of the actual encounter trains did not change --- all we change is the statistics of the disease onset time.
Please note further that the results for different versions of disease onset represent the extreme scenario where all disease onsets originate from external sources.

Comparing the distributions of $n_\text{inf}$ for fixed $T_\text{lat}$ (Fig.~\ref{fig:controls_infectious_encounters}a and Fig.~\ref{fig:controls_infectious_encounters}c each), we notice that those distributions that best resemble the shape of internal disease onset are those where external disease onset statistics incorporate the variability across individuals (external ii and iv).
This can be explained by the fact that also for internal disease onset more onsets occur for contact trains with more encounters, which in turn increases the probability of higher $n_\text{inf}$ and thereby also the mean $n_\text{inf}$.
It appears that for the overall shape of the distribution, as well as the leading order of its mean value, it is not necessary that disease onsets occur with the same temporal inhomogeneity as true encounters for the chosen infectious periods (this may change for very small infectious periods though).

Comparing further the results of specific disease onsets for different $T_\text{lat}$ (Fig.~\ref{fig:controls_infectious_encounters}b comparing solid vs opaque symbols), we notice that incorporating the temporal inhomogeneity into the disease onset (external disease ii and iii) seems relevant for the modulation that causes differences in $\overline{n}_\text{inf}$ due to latent periods.
This can be explained by reoccurring contact patterns: if infections are not more likely to occur during times of many contacts (as for i and iv), then differences in the number of potential infections are smeared out when averaging over multiple realizations.

Concluding, the results for the extreme scenarios of external infections fully support our conclusions from the main manuscript.

\begin{figure}[h]
    \centerline{\includegraphics[]{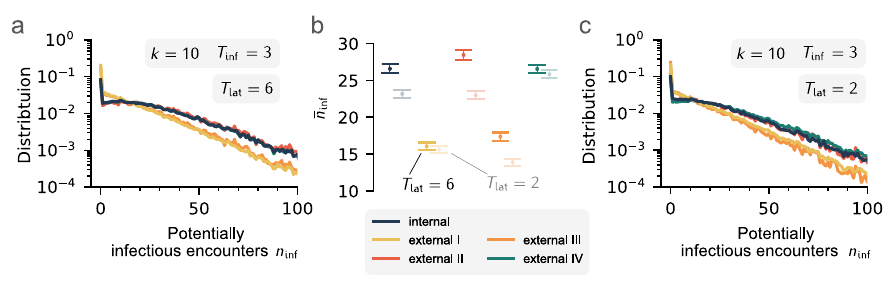}}
    \caption{
    \textbf{Infectious encounters for external infections.}
    In our analysis in the main manuscript, we preserved the temporal features by constraining disease onsets to available encounters.
    \textbf{a:}~Distribution of potentially infectious encounters for non-deterministic disease progression ($k=10$). 6 days latent period.
    \textbf{b:}~Comparison of the mean $\bar{n}_{\rm inf}$ between 6 days latent period (dark) and 2 days latent period (light).
    \textbf{c:}~Same as a), but 2 days latent period.
    }
    \label{fig:controls_infectious_encounters}
\end{figure}

\clearpage
\subsection{Impact of the distance threshold for considered contacts}
In the main manuscript, we filtered the physical proximity data to only include contacts with a minimal Bluetooth signal strength ($\text{RSSI}\gtrsim \SI{-80}{dBm}$) as a proxy for contact distances below $\SI{2}{m}$~\cite{sekara_strength_2014}.
This choice was motivated in the main manuscript and moreover corresponds to a value that optimizes the ratio between strong and
weak links~\cite{sekara_strength_2014}.
Here we show that our results are robust under changes in the cut-off signal strength which sets a scale, but does not alter the patterns observed.
In particular, we consider two cases of lower  ($\text{RSSI}\gtrsim\SI{-75}{dBm})$ and higher ($\text{RSSI}\gtrsim\SI{-95}{dBm})$ threshold.
As can be seen from Fig.~\ref{fig:sm_rssi}, this threshold changes the scale of the encounter rate (Fig.~\ref{fig:sm_rssi}a) but does not alter the resonance effect compared to random encounters (Fig.~\ref{fig:sm_rssi}b).

\begin{figure}[h]
    \centerline{\includegraphics[]{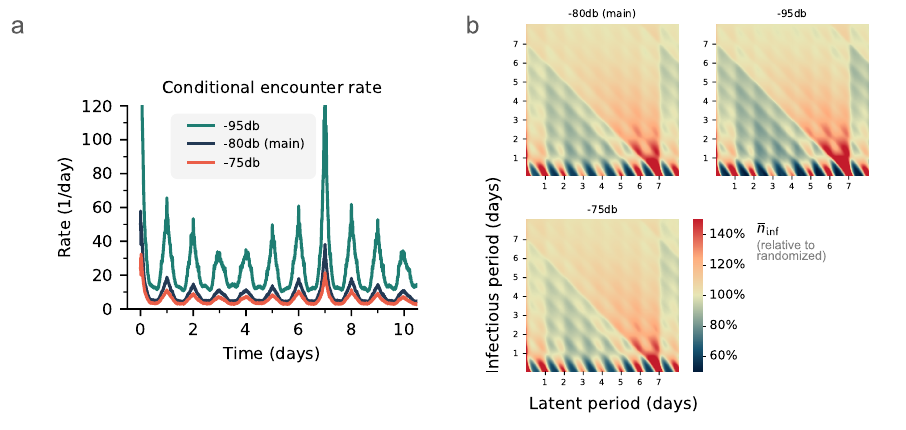}}
    \caption{
    \textbf{Effect of threshold distance}.
    To test which impact the distance at which a contact becomes potentially infectious has on our results,
we varied the $\text{RSSI}$ threshold in the preprocessing step.
A higher threshold ($\text{RSSI}\gtrsim\SI{-95}{dBm}$) corresponds to including more contacts (which previously would be considered too far away), thus resulting in a higher number of potentially infectious encounters, and (in absolute numbers) to a higher conditional encounter rate.
Vice versa, a lower threshold ($\text{RSSI}\gtrsim\SI{-75}{dBm}$, corresponding to about 1 meter distance) leads to a lower conditional encounter rate.
In both cases, the overall functional shape of the conditional encounter rate (featuring valleys at night, peaks at daytime and pronounced peaks at seven days) remains intact.
Consequently, also the number of potentially infectious encounters \textit{relative to randomized} remain mostly unaltered.
    }
    \label{fig:sm_rssi}
\end{figure}

\clearpage
\subsection{Impact of the contact duration for considered contacts}

In the main manuscript, we filtered contacts to only include those with a sufficient duration of at least $\SI{15}{minutes}$.
This choice was motivated in the main manuscript, and it is particularly suitable to avoid transient contacts that arise in passing. Still, different choices are possible.
Here, we show for shorter and longer minimum durations that this choice only affects the scale of the encounter rate but does not affect any results relating to the modulation of potentially infectious encounters (Fig.~\ref{fig:sm_mins}).


\begin{figure}[h]
    \centerline{\includegraphics[]{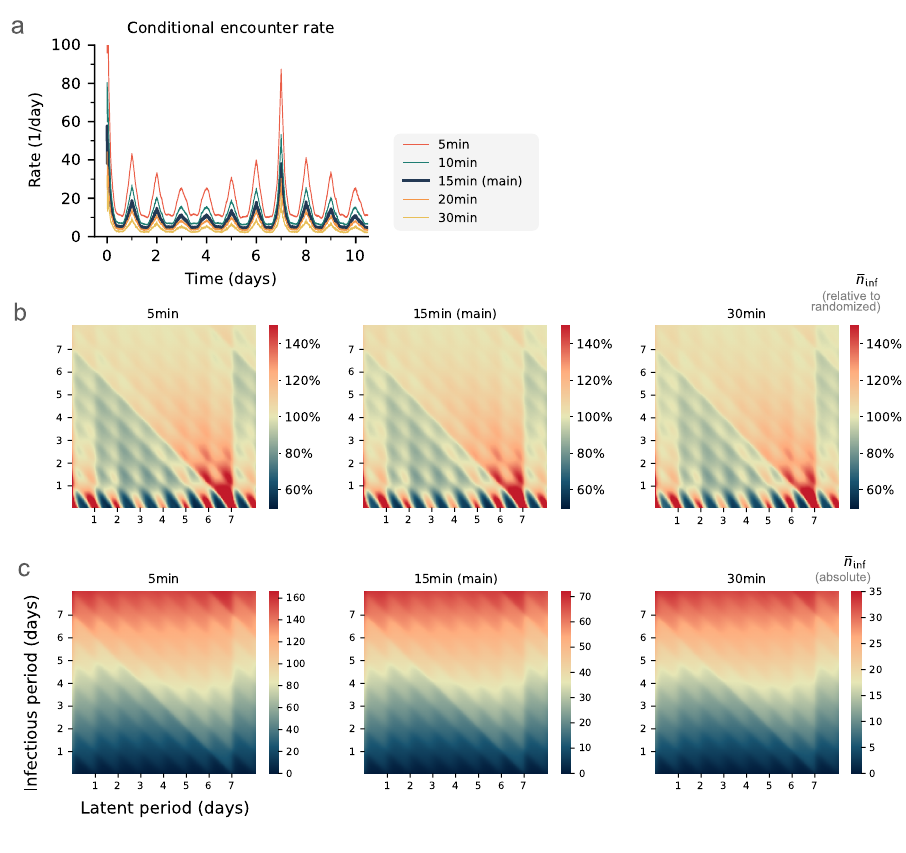}}
    \caption{
    \textbf{Effect of threshold duration}.
    To test which impact the minimum duration after which a contact becomes potentially infectious has on our results,
we varied the required duration for qualifying contacts during the preprocessing step. A shorter required duration (e.g. $\SI{5}{minutes}$ vs.~$\SI{15}{minutes}$ in the main manuscript) leads to more contacts qualifying and a higher conditional encounter rate. Vice versa, a longer required duration (e.g. $\SI{30}{minutes}$) leads to a lower conditional encounter rate.
In both cases, the overall functional form of the conditional encounter rate remains intact.
Again, although we see a higher absolute number of potentially infectious encounters, the change relative to randomized encounter trains is mostly unaltered.
    }
    \label{fig:sm_mins}
\end{figure}

\end{document}